\documentclass[apjl,a4paper,12pt,useAMS]{emulateapj}
\usepackage{amssymb}
\usepackage{graphicx}

\begin{document}
\title{A long-lived remnant neutron star after GW170817 inferred from its associated kilonova}
\author{Yun-Wei~Yu$^{1,2}$, Liang-Duan Liu$^{3,4}$, and Zi-Gao Dai$^{3,5}$}

\altaffiltext{1}{Institute of Astrophysics, Central China Normal
University, Wuhan 430079, China, {yuyw@mail.ccnu.edu.cn}}
\altaffiltext{2}{Key Laboratory of Quark and Lepton Physics (Central
China Normal University), Ministry of Education, Wuhan 430079,
China}\altaffiltext{3}{School of Astronomy and Space Science,
Nanjing University, Nanjing 210093, China, dzg@nju.edu.cn}\altaffiltext{4}{Department of Physics and Astronomy, University of Nevada, Las Vegas, NV 89154, USA
}\altaffiltext{5}{Key
Laboratory of Modern Astronomy and Astrophysics (Nanjing
University), Ministry of Education, China}

\begin{abstract}
The successful joint observation of the gravitational wave event
GW170817 and its multi-wavelength electromagnetic counterparts first
enables human to witness a definite merger event of two neutron
stars (NSs). This historical event confirms the origin
of short-duration gamma-ray bursts (GRBs), and in particular,
identifies the theoretically-predicted kilonova phenomenon that is
powered by radioactive decays of $r$-process heavy elements. However,
whether a long-lived remnant NS could be formed during this merger
event remains unknown, although such a central engine has been
suggested by afterglow observations of some short-duration GRBs. By
invoking this long-lived remnant NS, we here propose a model of
hybrid energy sources for the kilonova AT2017gfo associated with GW
170817. While the early emission of AT2017gfo is still powered
radioactively as usually suggested, its late emission is primarily
caused by delayed energy injection from the remnant NS. In our
model, only one single opacity is required and an intermediate value
of $\kappa\simeq0.97\,\rm cm^2g^{-1}$ is revealed, which could be naturally
provided by lanthanide-rich ejecta that is deeply ionized by the
emission from a wind of the NS. These self-consistent results indicate that
a long-lived remnant NS, which must own a very stiff equation of state, had been formed during the merger event of
GW170817. This provides a very stringent constraint on the strong interaction in nuclear-quark matter. It is further implied that such GW events could provide a probe of the early spin and magnetic evolutions of NSs, e.g., the burying of surface magnetic fields.
\end{abstract} 

\section{Introduction}

It has long been hypothesized that approximately half of the
elements heavier than iron in the Universe are synthesized via rapid
neutron-capture process ({\it r}-process) in the highly neutron-rich
outflows that come from mergers of a neutron star (NS) and a black hole (BH) or binary NSs \citep{Lattimer1974,Lattimer1976,Symbalisty1982}. The radioactive
decays of these {\it r}-process nuclei can effectively heat the merger ejecta and then cause nearly-isotropic kilonova emission, which provides a hopeful electromagnetic counterpart for the gravitational wave (GW) events due to the mergers.
Since the first
suggestion by \cite{Li1998} and the fundamental
development by \cite{Metzger2010}, the characteristics of kilonova emission have been
widely investigated
\citep{Kulkarni2005,Roberts2011,Kasen2013,Tanaka2013,Barnes2013,Yu2013,Metzger2014a,Grossman2014,Metzger2014b,Perego2014,Wanajo2014,Martin2015,Kasen2015,Li2016,Metzger2017b}.
In view of the ejecta mass of a few thousandth to hundredth solar masses, it is commonly predicted that the peak bolometric luminosity of a kilonova is around several times $10^{41}\rm erg~s^{-1}$ and its emission timescale is about a few days or somewhat longer.

Besides a nearly-isotropic
sub-relativistic ejecta, mergers of NS-NS and NS-BH binaries could also produce a pair of collimated relativistic jets
\citep{Rezzolla2011}, from which a gamma-ray burst (GRB) can be generated \citep{Paczynski1986, Eichler1989}. In other words, for an appropriate viewing angle, a kilonova is expected to accompany a short-duration GRB, both of which follow the preceding GW signal.
On the one hand, during the past few years, several kilonova candidates have been
tentatively identified from the optical-infrared emission in excess
of the afterglow emission of short GRBs, e.g., GRB 130603B
\citep{Tanvir2013,Berger2013}, 050709 \citep{Jin2016}, and 060614
\citep{Jin2015,Yang2015}, although the observational data are
always too scarce and ambiguous to draw a sufficiently solid
conclusion. On the other hand,  the widely-existing plateaus of X-ray afterglows and the extended soft gamma-ray emission of a remarkable number of
short GRBs strongly suggest that the remnant objects of these GRBs could be a rapidly rotating and highly magnetized massive NS, i.e., a millisecond magnetar \citep{Dai1998a, Dai1998b, Dai2004, Dai2006, Fan2006, Rowlinson2010,
Bucciantini2012, Rowlinson2013, Gompertz2013, Zhang2013,Lu2015,
Gompertz2015}. The equation of state (EOS) of such remnant NSs is likely to be very stiff, in view of their masses probably significantly higher than $\sim2M_{\odot}$. It was suggested that merger ejecta can be additionally powered by the spin-down of a remnant NS, which can substantially influence kilonova emission\footnote{Due to the
extra NS power, the luminosity of the ejecta emission can
in principle vary in a wide range, the upper bound of which can be much higher than $10^{41}\rm erg~s^{-1}$. Therefore,
\cite{Yu2013} suggested to term this emission
by a ``mergernova''. The mergernovae can have three sub-types: the NS-dominated, radioactivity-dominated, and hybrid ones on the focus of this paper.}
\citep{Kulkarni2005,Yu2013,Metzger2014a,Li2016}. This suggestion was supported by the simultaneous modeling of the multi-wavelength afterglows of GRB 130603B and its associated kilonova candidate \citep{Fan2013}.

Recently, a GW event (GW170817) was detected by
the advanced Laser Interferometer Gravitational-Wave Observatory
(LIGO) and Virgo Interferometer (Virgo) on 17 August 2017, which was, for the first time,
identified to come from a merger of two compact objects with typical masses of NSs
\citep{Abbott2017a}. This GW event was quickly found to be followed by GRB 170817A \citep{Abbott2017b,Goldstein2017,Savchenko2017,Zhang2017}, by GRB afterglow emission \citep{Troja2017,Margutti2017,Hallinan2017,Alexander2017,Lazzati2017,Lyman2018,DAvanzo2018}, and by an ultraviolet-optical-infrared (UVOIR) transient \citep{Coulter2017,Abbott2017c,Arcavi2017,Andreoni2017,Chornock2017,Covino2017,Cowperthwaite2017,Drout2017,Evans2017,Hu2017,Kasliwal2017,Kilpatrick2017,Nicholl2017,Pian2017,Shappee2017,Smartt2017,Soares-Santos2017,Tanvir2017}. On the one hand, these electromagnetic
counterparts of the GW event strongly indicated that the progenitor binary at least owns one NS. On the other hand,  the GW detection provided a smoking-gun for the long-hypothesized merger origin of short GRBs and confirmed the kilonova theory. The thermal spectra and the early light curve roughly
following a temporal behavior of $t^{-1.3}$ of the UVOIR transient, which is named AT2017gfo/SSS17a/DLT17ck (AT2017gfo hereafter), are well consistent with the theoretical predictions for kilonovae.

The discovery of kilonova AT2017gfo is undoubtedly one of the highlights of this milestone multi-messenger event. Different from previous kilonova candidates, the observations on AT2017gfo were comprehensive, timely, long lasting, multi-wavelength, and deep enough. Therefore, AT2017gfo provides an unprecedented opportunity to
observe the details of kilonova emission and, furthermore, to probe the ingredients of merger ejecta and the energy sources of the emission \citep{Kasen2017}. In this paper, we try to find a clue to the nature of the remnant object of the merger, by modeling the AT2017gfo emission. The paper is organized as follows. In Sections 2 and 3, the basic characteristics of AT2017gfo and our model are introduced, respectively. In Section 4, it is showed that the existence of a long-lived remnant NS is very helpful for understanding the AT2017gfo emission. Then, the possible properties of the remnant NS are discussed in Section 5. Finally, we give a summary and our conclusions in Section 6. 

\section{Kilonova AT2017gfo}
The comprehensive observations of AT2017gfo showed that the early and late phases of its emission cannot be explained by the
kilonova model with a single set of parameters, if only the radioactive power is invoked \citep{Cowperthwaite2017,Tanvir2017,Kilpatrick2017,Kasliwal2017,Villar2017,Shappee2017}. Therefore, it was widely
suggested that the AT2017gfo emission had evolved from a ``blue" emission stage to a ``red" stage, by considering that the merger ejecta could consist of different components of different opacities \citep{Metzger2014b,Kasen2015,Kasen2017}. Specifically, since a large number
of lanthanide elements can be formed in a highly neutron-rich
ejecta, the opacity of the ejecta at ultraviolet and optical wavelengths can be increased to be as high as
$\sim10-100~\rm cm^2g^{-1}$. This high opacity can effectively delay and
redden the ejecta emission, i.e., produce a so-called ``red"
emission \citep{Kasen2013,Tanaka2013,Barnes2013}. This emission is most likely produced by the tidal tail component of the merger ejecta in the
equatorial direction. On the contrary, in the polar direction where the ejecta material is
contributed by a squeezed outflow and a disk wind, the lanthanide synthesis can be suppressed, because the neutrino
irradiation from the remnant NS (if exists) can lead to an effective conversion of neutrons to
protons. Therefore, a relatively ``blue" and
fast-evolving emission can arise from this polar ejecta component, for a typical small opacity on the order $\sim0.1\rm ~cm^2g^{-1}$ \citep{Surman2008,Dessart2009,Wanajo2014,Perego2014,Martin2015}.

\cite{Villar2017} assembled all of
the available UVOIR data of AT2017gfo published and then mitigated the systematic
offsets between individual datasets. With the resulting pruned and
homogenized dataset, they fitted the UVOIR
light curves carefully and found that the AT2017gfo emission is dominated by a ``purple" component that corresponds to an
intermediate value of the opacity (i.e., $\sim3\rm ~cm^2g^{-1}$). On the contrary, the expected important ``red" emission component is relatively weak. In principle, such a ``purple" emission can be produced by the disk wind ejecta, if its lanthanide abundance is just appropriate due to an appropriate NS lifetime. Nevertheless, at the same time, the line of sight could still need to be fine tuned, because the disk wind ejecta is likely to be surrounded by the tidal tail ejecta of a higher velocity. Additionally, no matter in the ``blue+red" and the ``blue+purple+red" scenario, the total mass of the whole merger ejecta is always required to be higher than $\sim0.06M_{\odot}$, which is in fact
difficult to be produced by this merger event \citep{Shibata2017}.

In any case, the appearance of the ``non-red" emission component in AT2017gfo robustly indicates that the progenitor is a NS-NS binary and a massive remnant NS must be formed after GW170817.
The remaining question is how long this remnant NS can live. The answer to this question can provide a crucial constraint on the EOS of NSs, which is therefore highly concerned by the scientific community \citep{Abbott2017d,Ma2017,Pooley2017,Metzger2017b,Metzger2018,Murase2018}. Here, we argue that the difficulties of the traditional kilonova model in explaining the AT2017gfo emission indicate that the remnant NS of GW170817
can live for a very long time and even permanently. First of all, as suggested by \cite{Yu2013}, the spin-down of the long-lived remnant NS
can provide an extra power to the kilonova emission. Therefore, the required mass of the radioactive elements and thus of the ejecta can be somewhat smaller than that required by the single radioactive power model. Moreover, different from the
radioactive power distributing from the innermost to the outmost
ejecta, the NS power is all injected through the bottom of
the merger ejecta and, therefore, a longer diffusion time is needed.
In other words, the energy injection from the NS can naturally cause a
delayed emission component to dominate the late-time emission of the kilonova.
Second, the existence of the long-lived remnant NS is helpful for understanding the intermediate value of opacity. On the one hand, the neutrino emission from the NS can block the synthesis of Lanthanide elements, in particular, in the polar ejecta component. On the other hand, the high-energy emission from the NS wind could also reduce opacity by ionizing the lanthanide elements \citep{Metzger2014a}, if the elements have been synthesized previously, e.g., in the equatorial ejecta component. Therefore, in any case, the opacity of the
whole merger ejecta cannot be very high and then
the expected ``red" emission would
become ``purple" or even ``blue" no matter which direction the emission arises from.

\section{The hybrid-energy-source model}
Following the above considerations, we suggest that the nontrivial evolution of the  AT2017gfo emission could be caused by hybrid energy
sources including radioactivity and NS spin-down, as an alternative scenario
to the widely-considered ``radioactive power + multiple opacity"
model.

Our calculations of kilonova emission are implemented in a simplified
radiation transfer model given by \cite{Kasen2010} and \cite{Metzger2017a}, which can well
approximate the radiation transfer and reveal the influence of mass
distribution. Specifically, the merger ejecta is considered to consist of an ensemble of a series of mass
layers. The mass of each layer is determined by the density
distribution, which is taken as a power law as
\citep{Nagakura2014}
\begin{equation}
\rho_{\rm ej}(R,t)={(3-\delta)M_{\rm ej}\over 4\pi
R_{\max}^3}\left[\left({R_{\min}\over
R_{\max}}\right)^{3-\delta}-1\right]^{-1}\left({R\over
R_{\max}}\right)^{-\delta},
\end{equation}
where $M_{\rm ej}$ is the total mass and $R_{\max}$ and $R_{\min}$
are the radii of the head and bottom of the ejecta, respectively.
Since the internal energy involved in AT2017gfo is much smaller
than the kinetic energy of the ejecta, the dynamical evolution of
each layer is neglected. In other words, each layer is considered to
evolve nearly independently with a constant velocity $v_i$.
Therefore, the maximum and minimum radii of the ejecta for a time
$t$ can be calculated by $R_{\max}(t)=v_{\max}t$ and
$R_{\min}(t)=v_{\min}t$ by introducing the maximum and minimum
velocities. Here we denote the mass layers by the subscript $i=1,2,\cdot
\cdot\cdot,n$, where $i=1$ and $n$ represent the bottom and the head
layers, respectively. The thermal energy $E_{i}$ of the $i$-th layer
evolves according to
 \begin{equation}
{dE_{i}\over dt}=\xi L_{\rm md}+m_{i}\dot{q}_{{\rm r},i}\eta_{\rm
th}-{E_{i}\over R_{i}}{dR_{i}\over dt}-L_{i} ,{\rm ~for }~i=1,
\label{Econ1}
\end{equation}
or
\begin{equation}
{dE_{i}\over dt}=m_{i}\dot{q}_{{\rm r},i}\eta_{{\rm th}}-{E_{i}\over
R_{i}}{dR_{i}\over dt}-L_{i}, {\rm ~for }~i>1,
\end{equation}
where $L_{\rm md}$ is the power carried by a NS wind that can be estimated by the magnetic dipole (MD) radiation luminosity of the NS, $\xi$ is the absorption fraction of this wind energy by the merger ejecta, $\dot{q}_{\rm r}$ is the radioactive power per unit mass, $\eta_{\rm th}$ is the thermalization efficiency of the radioactive power, $m_{i}$ and $R_{i}$ are the mass and radius
of the layer, and $L_i$ is the observed luminosity contributed by
this layer. The NS power comes from the
central cavity of the ejecta and penetrates into the ejecta through
its bottom. Therefore, it is simply considered that the
NS power term only appears in Equation (\ref{Econ1}) for
$i=1$, which means that the energy is primarily absorbed by the
bottom material.

The temporal evolution of the luminosity of MD radiation of a NS can be expressed as
\begin{equation}
L_{\rm md}(t)=L_{\rm md}(0)\left(1+{t\over t_{\rm
sd}}\right)^{-\alpha}\label{Lsdt},
\end{equation}
with an initial value of
 \begin{eqnarray}
L_{\rm md}(0)&=&{B_{\rm p}^2R_{\rm s}^6\over{6c^3}}\left({2\pi\over
P_{\rm i}}\right)^4\nonumber\\
&=&9.6\times10^{48}R_{\rm s,6}^{6}B_{\rm
p,15}^{2}P_{\rm i,-3}^{-4}\rm erg~s^{-1},\label{luminosity}
\end{eqnarray}
where $R_{\rm s,6}=R_{\rm s}/10^6$cm, $B_{\rm p,15}=B_{\rm
p}/10^{15}$G, and $P_{\rm i,-3}=P_{\rm i}/1$ms are the radius, the surface
polar magnetic field, and the initial spin period of the
NS, respectively, and $c$ is the speed of light.
The spin-down timescale $t_{\rm sd}$ in Equation (\ref{Lsdt}) can in principle be determined
by either MD radiation or secular GW radiation of the NS, corresponding
to a decay index of $\alpha=2$ or 1, respectively.
The radioactive power per unit mass can be written as \citep{Korobkin2012}
\begin{equation}
\dot{q}_{\rm r}=4\times10^{18}\left[{1\over2}-{1\over \pi}{\rm
arctan}\left({{t-t_0}\over \sigma}\right)\right]^{1.3}\rm
erg~s^{-1}g^{-1}\label{eq:radioa}
\end{equation}
with $t_0=1.3$\,s and $\sigma=0.11$\,s. Its thermalization efficiency
reads \citep{Barnes2016,Metzger2017b}
\begin{equation}
\eta_{\rm th}=0.36\left[\exp(-0.56 t_{\rm day})+{\ln (1+0.34t_{\rm
day}^{0.74})\over 0.34t_{\rm day}^{0.74}}\right],\label{eq:effic}
\end{equation}
where $t_{\rm day}=t/1\rm day$.

\begin{figure}
\includegraphics[width=0.5\textwidth]{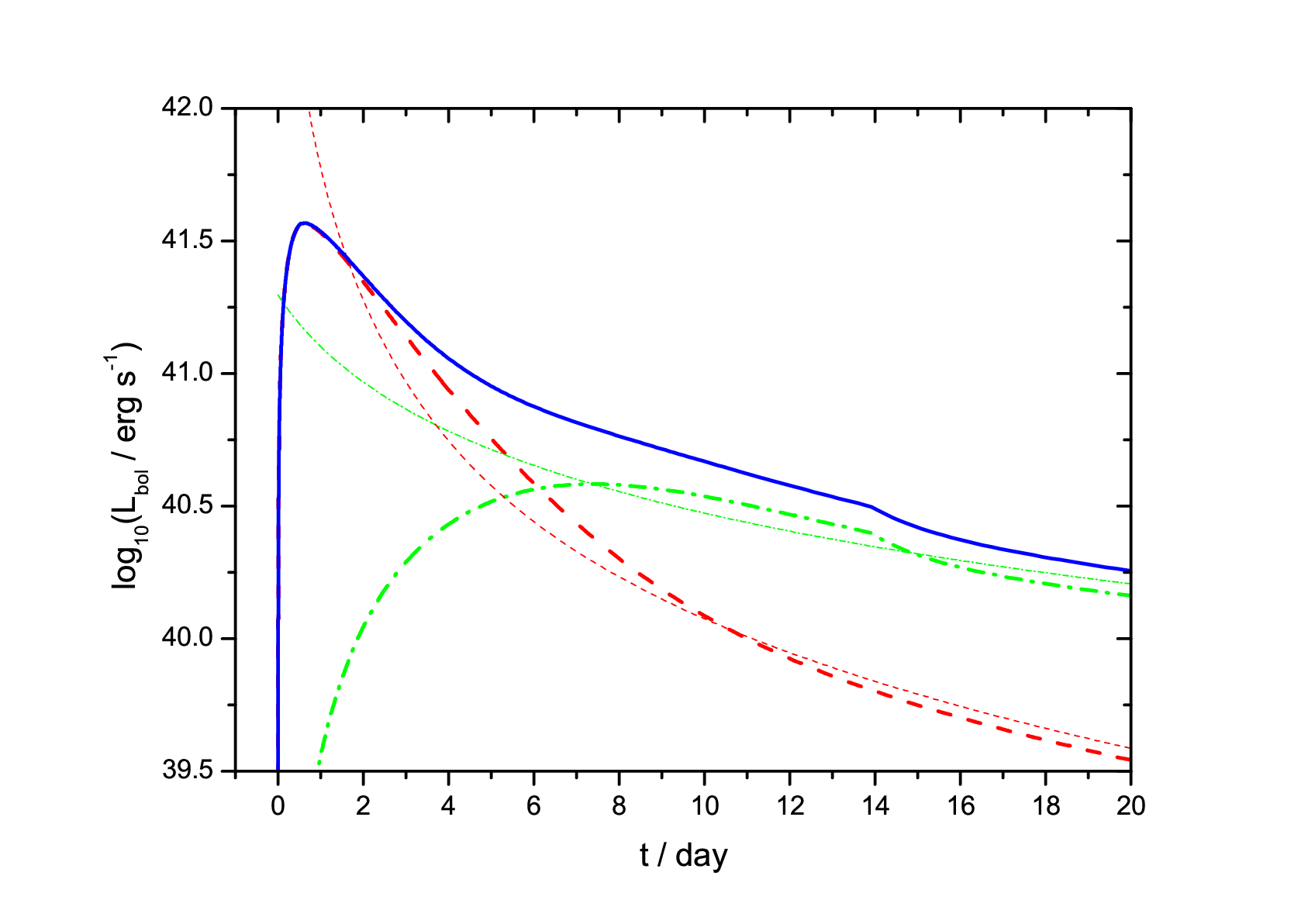}
\caption{Bolometric light curve of a kilonova of hybrid energy
sources. The thin dashed and dash-doted lines represent the heating
power of radioactivity and of NS spin-down, respectively. The thick
dashed and dashed-dotted lines are bolometric light curves powered
by the corresponding single energy source. The solid line is the
result of the combination of the two energy sources. The model parameters are taken as the central values presented in Table \ref{table1}.}\label{bolometric}
\end{figure}

The observed luminosity contributed by the $i$-th layer can
be estimated by
 \begin{equation}
L_{i}={E_{i}\over \max[t_{{\rm d},i},t_{{\rm lc},i}]},
\end{equation}
where $t_{{\rm d},i}$
represents the radiation diffusion timescale, during which the heat
escapes from the whole ejecta, and $t_{{\rm lc},i}=R_{i}/c$ the time limit given by the light
crossing that ensures the causality. Here, the radiation diffusion
timescale of the $i$-th layer is given by
 \begin{equation}
t_{{\rm d},i}={3\kappa \over 4 \pi R_{i}c}
{\sum\limits_{i'=i}^{n}}m_{i'},
\end{equation}
where the effects of the shells external to the $i$-th layer are
all taken into account by summarizing the layer masses from $i$ to
$n$. In our calculations, a uniform opacity will be taken.
Finally, the total bolometric luminosity of the merger ejecta can be obtained by summarizing the contributions of all layers:
 \begin{equation}
L_{\rm bol}={\sum\limits_{i=1}^{n}}L_{i}.
\end{equation}
In Figure \ref{bolometric}, we present an example bolometric light curve of a kilonova powered by hybrid energy sources. As shown, for given parameter values, the energy inejction from the remnant NS can gradually dominate the late kilonova emission, while the emission in the first few days is controlled by the radioactive power.

\begin{figure}
\centering\includegraphics[width=0.5\textwidth]{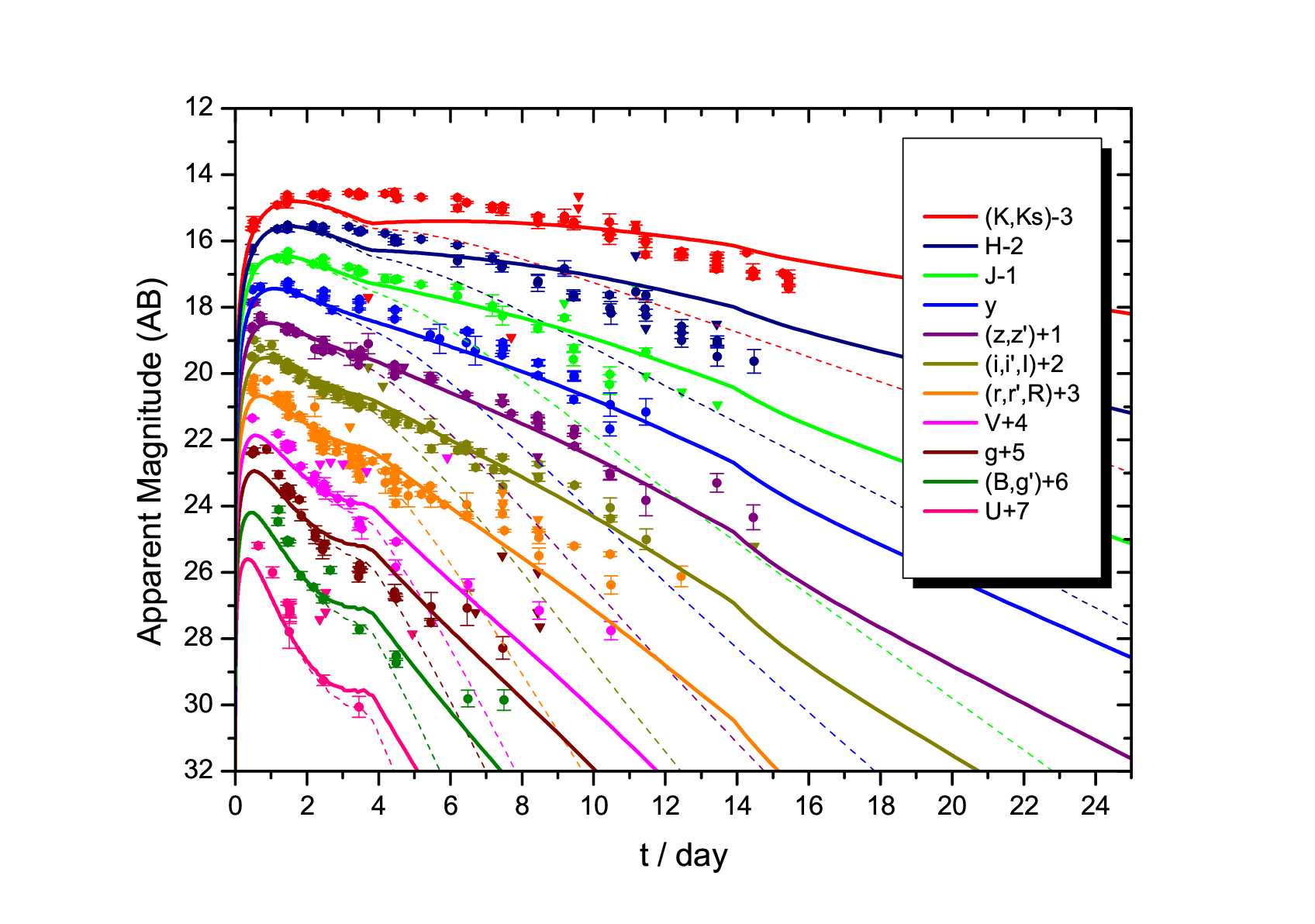}
\caption{UVOIR light curves of kilonova AT2017gfo. The observational data (circles) or limits (triangles)
are taken from Villar et al. (2017). The solid lines are given by our hybrid-energy-source model, while
the dashed lines represent the results only with the radioactive
power, where the model parameters are taken as the central values presented in Table \ref{table1}. Black-body spectra are assumed all the time.
}\label{filters}
\end{figure}

The observational
UVOIR light curves of AT2017gfo are given in Figure \ref{filters}, which are taken from
\cite{Villar2017} and presented in different filters. In order to fit these monochromatic light curves, a frequency-dependent radiative transfer in principle needs to be taken into account to get the emission spectra. In this paper, for simplicity, we assume the spectra to always be a black body. Then, the effective temperature of
the kilonova emission can be determined by
\begin{equation}
T_{\rm eff}=\left({L_{\rm bol}\over4\pi\sigma_{\rm SB} R_{\rm ph}^2}\right)^{1/4},
\end{equation}
where $\sigma_{\rm SB}$ is the Stephan-Boltzmann constant and the photospheric radius
$R_{\rm ph}$ is defined as that of the mass layer beyond which the
optical depth equals unity. After the total optical depth of the ejecta
is smaller than unity, we fix $R_{\rm ph}$ to $R_{\min}$. Then, the
flux density of the kilonova emission at photon frequency $\nu$ can
be given by
 \begin{equation}
F_{\nu}(t)={2\pi h\nu^3\over c^2}{1\over \exp(h\nu/kT_{\rm
eff})-1}{R_{\rm ph}^2\over D^2},
\end{equation}
where $h$ is the Planck constant, $k$ is the Boltzmann constant, $D=40$ Mpc
is the distance of the source. Finally, we determine the
monochromatic AB magnitude by $M_{\nu}=-2.5\log_{10}(F_{\nu}/3631\rm
Jy)$.

\begin{table}[tbph]
\caption{Priors and posteriors of the model parameters} \label{table1}
\begin{center}
\begin{tabular}{llll}
\hline \hline
Parameter   & Prior  & Allowed Range  & Constraint
\\ \hline
 $\xi L_{\rm md}(0)/10^{41} \rm{erg \ s}^{-1}$   & Flat & [0.1, 100]   & $1.98 \pm0.03$\\
 $t_{\rm sd}/10^{5}$s                            & Flat & [0.01, 10]   & $1.53^{+0.02}_{-0.03}$\\
 $M_{\rm{ej}}/0.01M_{\odot}$                     & Flat & [0.1, 10]    & $2.98\pm0.02$           \\
 $\kappa/ \rm{g \ cm}^{-2}$                      & Flat & [0.1, 10]  & $0.97^{+0.01}_{-0.02}$\\
 $v_{\rm{min}}/c$                                & Flat & [0.01, 0.15] & $0.10^{+0.02}_{-0.01}$         \\
 $v_{\rm{max}}/c$                                & Flat & [0.18, 0.40]  & $0.40^{+0.00}_{-0.01}$\tablenotemark{a} \\
 $\delta$                                        & Flat & [1.0, 3.0]   & $1.46^{+0.04}_{-0.05}$\\
 \hline
\end{tabular}%
\end{center}
\par
Note: $^{\rm a}$Parameter reaches the limit of the allowed range.
\end{table}

\section{Analysis and fitting}
Before specific fittings, some order of magnitude analyses can be made on the primary model parameters. On the one hand, the early emission of AT2017gfo can well be described by black-body spectra. Then it can easily be revealed that the bolometric luminosity of AT2017gfo reached the peak of $L_{\rm bol,p}\sim5\times10^{41}\rm erg~s^{-1}$ at around $t_{\rm p}\sim5000\,$s after GW170817. In our model, this early emission is primarily powered by radioactivities and nearly unaffected by the NS power. So, according to the so-called Arnett Law \citep{Arnett1980}, the mass of the ejecta can roughly be estimated by $M_{\rm ej}\sim L_{\rm bol, p}/[\eta_{\rm th}(t_{\rm p})\dot{q}_{\rm r}(t_{\rm p})]$, which is about several times $0.01\rm M_{\odot}$. At the same time, the opacity of the ejecta is found to be not very much higher than the order of $0.1\rm cm^2g^{-1}$, which can be derived from the analytical estimation of the peak emission time as $t_{\rm p}\sim(\kappa M_{\rm ej}/4\pi v_{\rm ej}c)^{1/2}$. This indicates that the AT2017gfo emission is ``blue" or ``purple", including the late emission that is considered to be contributed by the same ejecta in our model.

On the other hand, from the late emission of AT2017gfo, it can be inferred that the absorbed power by the ejecta from the wind of the remnant NS cannot be much higher or lower than the kilonova luminosity and meanwhile the spin-down timescale of the NS could be comparable to a few days. If the NS is braked by MD radiation, the spin-down timescale can be expressed as
 \begin{equation}
t_{\rm sd,md}={{3Ic^3}\over {B_{\rm p}^2R_{\rm
s}^6}}\left({2\pi\over P_{\rm
i}}\right)^{-2}=2\times10^3I_{45}R_{\rm s,6}^{-6}B_{\rm
p,15}^{-2}P_{\rm i,-3}^2\rm s.\label{tsdmd}
\end{equation}
Combining the estimations of $\xi
L_{\rm md}(0)\sim10^{41}\rm erg~s^{-1}$ and $t_{\rm sd}\sim3$ day with Equations (\ref{luminosity}) and (\ref{tsdmd}), an extremely high magnetic field strength $B_{\rm
p}=7.8\times10^{16}\xi^{1/2}I_{45}R_6^{-3}\rm G$
and a very long initial spin period $P_{\rm
i}=870~\xi^{1/2}I_{45}^{1/2}\rm ms$ can be derived. Both of these two parameter values seem somewhat unrealistic. In particular, the spin period is too long to be consistent with the Keplerian period (i.e., $\sim 1$ ms)
that a post-merger NS is considered to rotate at. The parameter $\xi$, which represents the fraction of the NS power absorbed by the merger ejecta, could be much smaller than unity, based on the following two reasons. (1) The overwhelming majority of the energy of the NS wind could be collimated into a small cone that points to the GRB jet, and (2) a remarkable fraction of energy could be reflected back into the wind when the wind emission encounters with the bottom of the merger ejecta \citep{Metzger2014a}. In any case, even for a very small $\xi\sim 0.01$, the obtained spin period would still be much
longer than $\sim 10$ ms, which is unacceptable for a post-merger NS. In summary, the MD-braking
scenario is disfavored. Therefore, we prefer to suggest that the remnant NS of GW170817 is primarily braked by secular GW radiation (i.e., $\alpha=1$), at least, during the kilonova emission period.


\begin{figure*}
\centering
\includegraphics[width=0.8\textwidth]{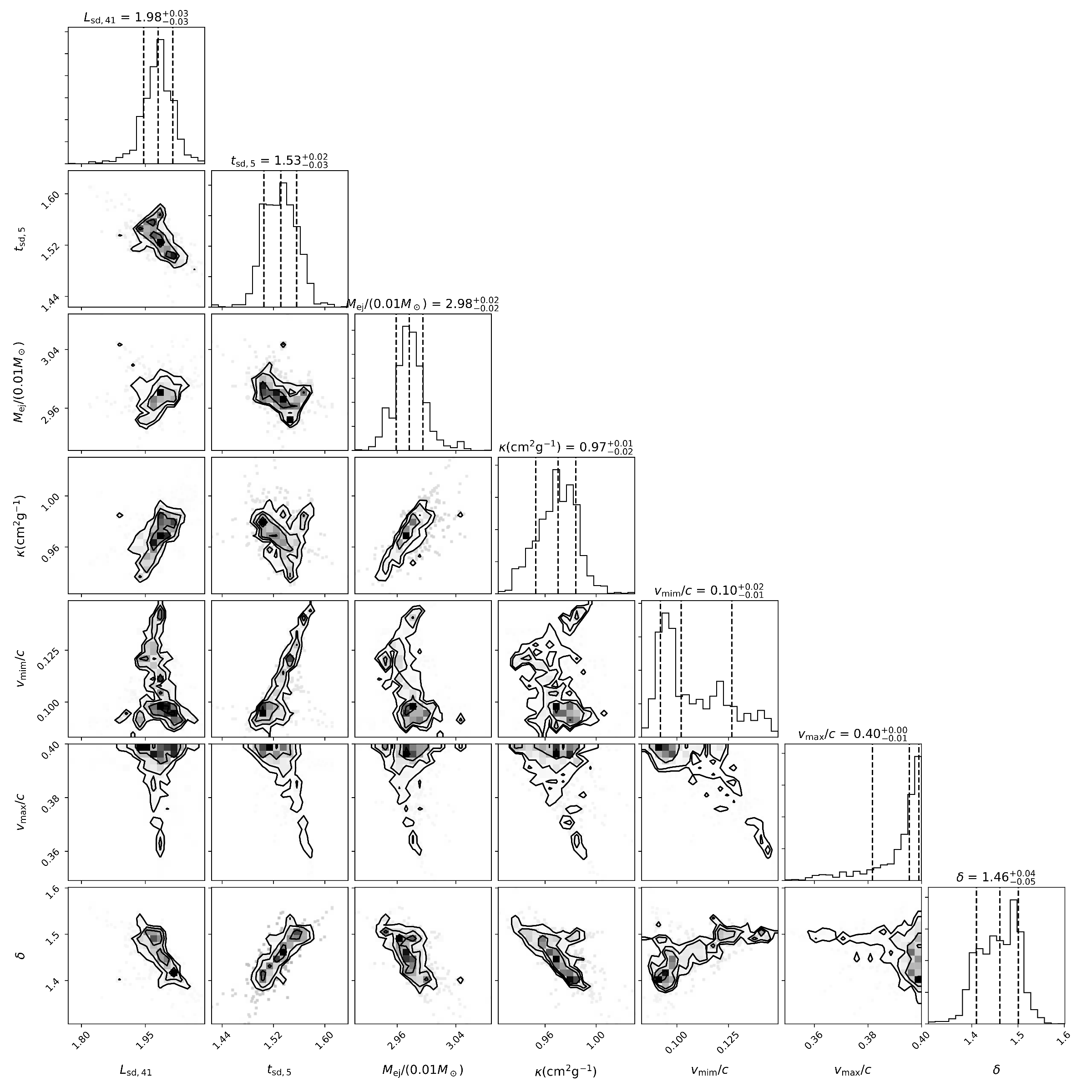}
\caption{Corner plot of the parameters for the fittings of the LCs of AT2017gfo. Medians and 1$\sigma$ ranges are labeled.}\label{MCMC}
\end{figure*}

To be specific, there are seven free parameters in our model, as listed in Table \ref{table1}, including five ejecta parameters (i.e., $M_{\rm ej}$, $\kappa$, $v_{\min}$, $v_{\max}$, and $\delta$) and two NS parameters (i.e., $\xi L_{\rm md}(0)$ and $t_{\rm sd}$ for $\alpha=1$).
  When we fit the observational UVOIR light curves of AT2017gfo, a Markov Chain Monte Carlo (MCMC) method is adopted to minimize the $\chi^2$ of the fitting and then to constrain the parameter ranges. The priors of the parameters are taken to be flat in the linear space of allowed ranges as given in Table \ref{table1}.
 The {\tt emcee} code \citep{For2013} is used for our MCMC fitting. We adopt a ``walkers'' number of $n_{\rm walkers} =50$ and $4000$ steps of those chains, where the first 100 iterations is used to burn in the ensemble. The code is run in parallel by using 16 nodes with a duration of 150 hours.  As a result, the constrained contours and the model parameters at the $1\sigma$ confidence level are presented in Figure \ref{MCMC} and Table \ref{table1}, respectively. 

 For the best-fitting parameter values, we present the theoretical light curves in Figure \ref{filters} in comparison with the observational data, which shows that our model provides a plausible explanation for the AT2017gfo emission. Some deviations between the theoretical curves and the data in some individual filters primarily arise from the black body
simplification of the spectra.

As a result, a relatively normal ejecta mass of $0.03\pm0.002M_{\odot}$ and
an intermediate opacity $\kappa\simeq0.97^{+0.01}_{-0.02}\,\rm cm^2g^{-1}$ are revealed by our fitting, while the velocity of the
ejecta of $\sim(0.1-0.4)c$ is well consistent with those found by previous works
\citep{Kilpatrick2017,Kasliwal2017,Smartt2017,Arcavi2017,Cowperthwaite2017,Villar2017}. These results are basically in agreement with the order of magnitude analysis given in the beginning of this section. To be specific, the obtained ejecta mass, which is about half of that found by \cite{Villar2017}, is relatively easy to be accounted for by NS-NS mergers.
And the intermediate opacity indicates that the AT2017gfo emission is somewhat ``purple", as also found by \cite{Villar2017}. We suggest that the ``purple" AT2017gfo emission is produced by an ejecta lanthanide-rich but deeply ionized, which is probably the tidal tail ejecta being irradiated by the high-energy emission from the NS wind. This explanation of AT2017gfo is insensitive to the opening angle of the polar ejecta component and
the viewing angle of observers, because the tidal tail ejecta is
observable in all directions.
In contrast, the
popular multiple-opacity model could usually require a particular ejecta structure and a fine-tuning viewing angle.

The NS parameters are constrained to be $t_{\rm sd}=1.53^{+0.02}_{-0.03}\times10^{5}$s and $\xi L_{\rm
md}(0)=1.98\pm0.03 \times10^{41}\rm erg~s^{-1}$, as expected. Because of the GW braking, the spin-down timescale of the remnant NS can be
calculated by
 \begin{equation}
t_{\rm sd,gw}={5P_{\rm
i}^4c^5\over{2048\pi^4GI\epsilon^2}}=9.1\times10^{5}\epsilon_{-4}^{-2}I_{45}^{-1}P_{\rm
i,-3}^4\rm s,\label{tsdgw}
\end{equation}
where $G$ is the gravitational constant, $I_{45}=I/10^{45}\rm
g~cm^2$ is the moment of inertia, and $\epsilon$ is the ellipticity
of the NS.
Then, the combination of the observationally constrained parameter values with Equations (\ref{tsdgw}) and (\ref{luminosity}) can yield
 \begin{equation}
B_{\rm p}=1.4\times10^{11}\xi^{-1/2}R_6^{-3}P_{\rm i,-3}^2\rm G,\label{B_GW}
\end{equation}
and
\begin{equation}
\epsilon=3.4\times10^{-4}I_{45}^{-1/2}P_{\rm i,-3}^2.\label{eps_GW}
\end{equation}
This surface magnetic field $B_{\rm p}$, which is consistent with that inferred from the X-ray flares observed in some short GRBs \citep{Dai2006}, is somewhat lower than the standard values for Galactic radio pulsars (i.e., $10^{12}$G), even though a small value of $\xi$ is taken. On the contrary, the relatively high ellipticity $\epsilon\sim10^{-4}-10^{-3}$ implies that the internal and probably toroidal magnetic fields of
the NS is ultra-high, if the ellipticity is induced by
the internal fields as $\epsilon\approx10^{-4}(B_{\rm int}/10^{16}\rm G)^2$ \citep{Cutler2002}. This high discrepancy between the surface and internal magnetic fields indicates that the MD radiation of the NS must be significantly suppressed due to some processes as discussed in the next section.

\section{Discussions on the remnant NS}

\subsection{EOSs}
From the LIGO observation, the chirp mass of the progenitor binary of GW170817 can be derived to $M_{\rm c}=1.188^{+0.004}_{-0.002}\rm ~M_{\odot}$, which constrains the individual masses of the component NSs to be in the range of $1.17-1.6\rm~ M_{\odot}$ by assuming low spins for the NSs \citep{Abbott2017b}.
The total gravitational mass of the binary can further be estimated to be about $2.74\rm ~M_{\odot}$ \citep{Abbott2017b}.  Then, after the GW chirp and mass ejection, the remnant NS could own a gravitational mass of about $M_{\rm RNS}\sim2.6\rm~ M_{\odot}$ \citep{Ai2018,Banik2017}.

As widely suggested, the remnant NS could be supported by its extremely rapid rotation, i.e., it could be a supramassive NS. Then, as the spin-down of the NS, it would quickly collapse into a BH, which is considered to be responsible for the launching of the GRB jets. Following this consideration, the maximum mass of a non-rotating NS $M_{\rm TOV}$ is required to be not higher than $M_{\rm RNS}/\lambda=2.16\rm M_{\odot}$ \cite[e.g.,][]{Margalit2017,Ruiz2018,Rezzolla2018,Shibata2017,Banik2017}, where $\lambda\approx1.2$ is the ratio of the maximum mass of a uniformly rotating NS to that of a non-rotating star.

On the contrary, in this paper, it is found that the kilonova emission of AT2017gfo requires the remnant NS to be stable persistently, at least, for about 20 days after the merger. In this case, therefore, the maximum mass of a non-rotating NS should satisfy $M_{\rm TOV}>M_{\rm RNS}$. Such a high maximum mass provides a very stringent constraint on the EOS of the remnant NS. Nevertheless, several EOSs can still survive this ordeal, e.g., the MS1 and SHT EOSs in \cite{Piro2017}, the NL3$\omega\rho$ EOS in \cite{Zhu2018}, the PMQS3 EOS in \cite{LiA2017}, and the MS1, MS1b, eosL, GS1, Heb4, Heb5, Heb6, LS375, and NL3 EOSs in \cite{Gao2017}. All of these EOSs can give a static maximum mass larger than $\sim2.6\rm ~M_{\odot}$, which indicates that our argument of a long-lived remnant NS after GW170817 is allowed in principle. Such a NS has become more expectable since the discovery of two binary NS systems with total gravitational masses as low as $\sim2.5M_\odot$ \citep{Martinez2017,Stovall2018}. Following this argument, \cite{Geng2018} fitted the late-time broadband afterglow of GW170817 very well.

In future works, it will be necessary and meaningful to investigate and test the other astrophysical consequences of these surviving EOSs, e.g., the merger processes as numerically studied previously \cite[e.g.,][]{Kastaun2015} and the tidal deformability\footnote{The LIGO observation of GW170817 provided a constraint on the tidal deformability of the pre-merger NSs as $\Lambda\leq 800$ in low-spin case or  $\Lambda\leq 1400$ in high-spin case, according to which the NS EOSs can be constrained \citep[e.g.,][]{Annala2018,Zhou2018,Nandi2017,Zhu2018}. However, since the pre- and post-merger NSs are actually produced through completely different processes (i.e., core-collapse vs merger), they could in principle be constituted by very different matter states. In this case, the $\Lambda-$constraint on the pre-merger NSs may not be extended to the post-merger NS. } of the corresponding NSs.

\subsection{Magnetic fields}
In order to account for the secular GW radiation, the internal magnetic fields of the remnant NS is required to be not much lower than $\sim10^{16}$ G, which probably determines a surface magnetic field of the order $\sim 10^{14}-10^{15}$ G. However, as restricted by the luminosity of AT2017gfo, the effective surface dipolar magnetic field of the remnant NS can only be of the order $\sim10^{11}-10^{12}$ G at the kilonova timescale. This discrepancy in the surface magnetic field indicates that the surface field could be hidden\footnote{Here the surface magnetic field is considered to be just hidden, but not be annihilated or dissipated. Therefore, no extra energy release is expected to happen during this period.} significantly by some reasons at a certain time. Before that moment, the remnant NS exhibits as a normal magnetar, which can produce a luminous wind emission of a luminosity of $\sim10^{48}\rm erg~s^{-1}$ and effectively ionize the merger ejecta. At the same time, the wind emission can also contribute a bright internal plateau component to the GRB afterglow emission at the jet direction, which was unfortunately missed for GRB 170817A because of the off-axis observations \citep{Ioka2017,Xiao2017,Granot2017,Salafia2017,Murguia2017,Lazzati2017}.

Here, we would like to mention an observational fact that an extremely steep decay following a bright plateau was usually observed in the afterglow light curves of many regular short GRBs. The post-plateau steep decay was previously explained as the collapse of a massive NS into a BH
\citep{Troja2007,Rowlinson2010,Rowlinson2013}.  However, as found in this paper, the ``internal plateau + steep decay" structure of GRB afterglows is probably caused by an abrupt suppression of the MD radiation of a remnant NS, rather than by the collapse of the NS.
Following this consideration, the time when the MD radiation is suppressed can be inferred from the timescales of the post-plateau steep decay of $\sim10^3-10^4$\,s. The zero time adopted in our fitting calculations is actually defined by this time.

In our Galaxy, several young, isolated, radio-quiet NSs, locating close to the center of $\sim$kyr supernova remnants, were discovered to have relatively low surface magnetic field. These NSs called central compact objects (CCOs) were widely suggested to be ``antimagnetars" whose strong magnetic field on the stellar surface is compressed and buried into the NS crust by hypercritical fallback accretion onto the NS \citep{Muslimov1995,Young1995,Geppert1999,Shabaltas2012,Torres-Forne2016}.  Therefore, it could be reasonable to consider that the remnant NS of GW170817 was also screened by fallback material, which hid the surface magnetic field of the NS and thus suppress the MD radiation.

\begin{figure*}
\centering\includegraphics[width=0.5\textwidth]{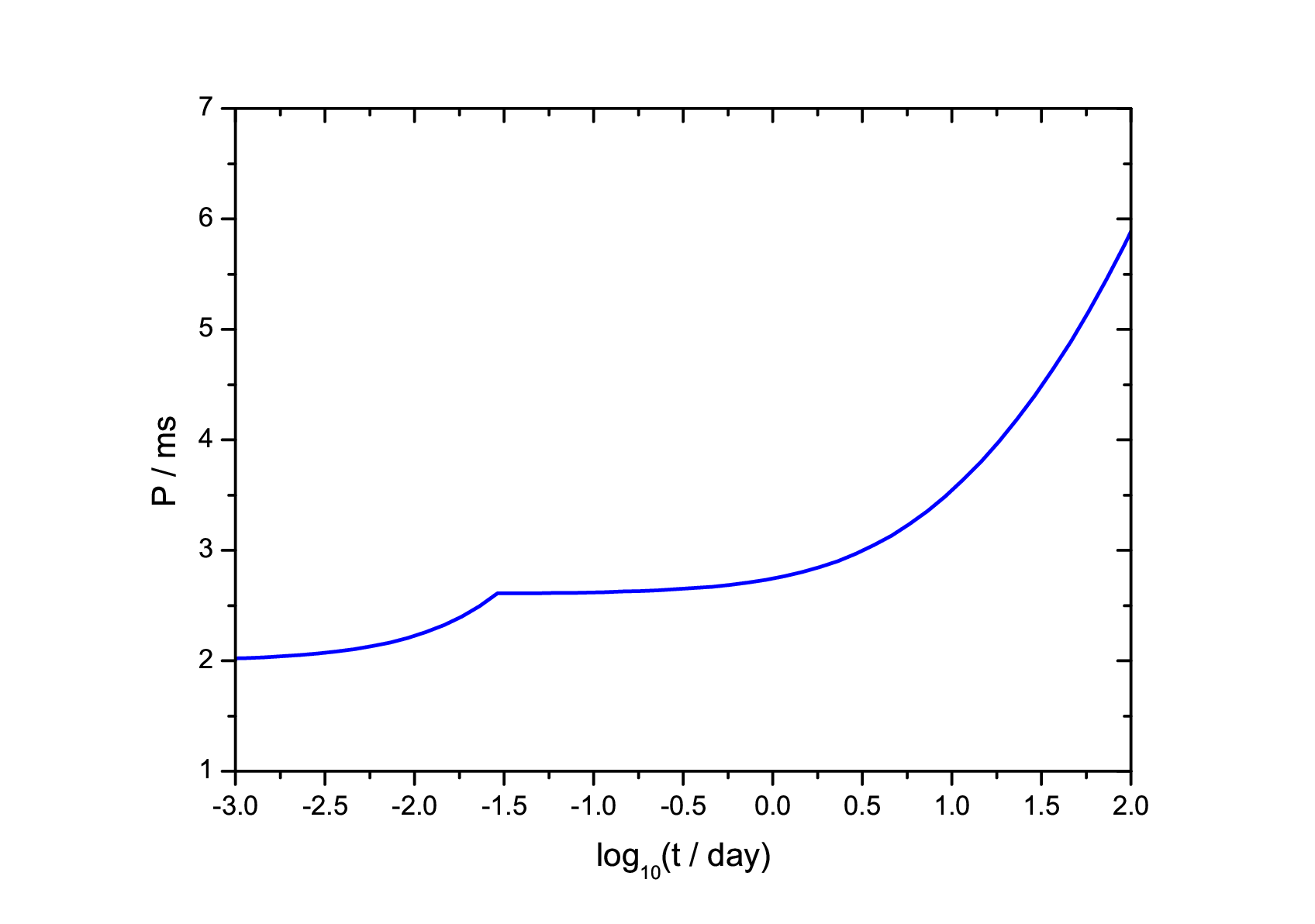}\includegraphics[width=0.5\textwidth]{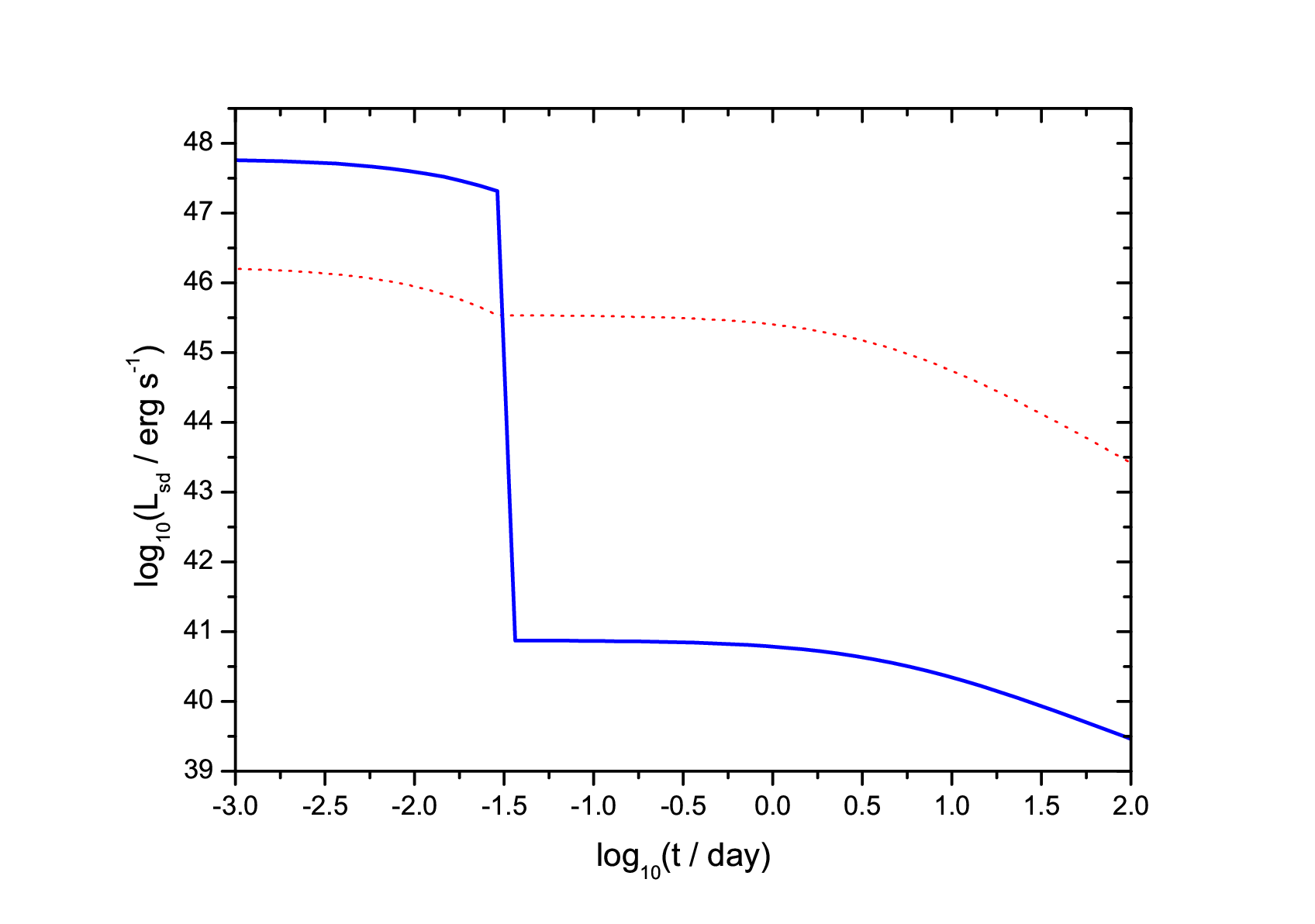}
\caption{An illustration of the evolution of spin period (left) and spin-down luminosity (right) of a post-merger NS, where a sharp decline of the surface magnetic field is considered to happen at the time of 3000 s. The dotted line gives the GW luminosity for a comparison.}\label{NSevolution}
\vspace{5mm}
\end{figure*}

In the case of NS-NS mergers, the fallback material could come from the slowest tail of the merger ejecta, the velocity of which is not much higher than the escaping velocity of the system.
Reverse shocks arising from some density discontinuities in the merger ejecta could play an important role in causing the falling of the material, as discussed in the supernova cases. It is difficult to estimate the starting time of the fallback accretion and the amount of the fallback material without numerical simulations. In any case, the timescale during which the majority of material is accreted by the NS can still be estimated by the free-fall time as
\begin{equation}
t_{\rm acc}\sim{1\over2}\left({r_{\min}^3\over GM}\right)^{1/2}=870{\rm s}\left({M\over 2.5M_{\odot}}\right)^{-1/2}\left({r_{\min}\over 10^{11}\rm cm}\right)^{3/2},
\end{equation}
where $r_{\min}$ is the radius of the base of ejecta when the fallback accretion starts. Therefore, the burying time of the surface magnetic field of the NS can be settled according to this timescale, even though the fallback accretion continues as a temporal behavior of $t^{-5/3}$. Here, one may worry about whether the amount of the fallback material is large enough to screen the whole NS. We would like to point out that what we directly derived from the AT2017gfo observations is just the energy release from the remnant NS. Therefore, in principle, the amount of the fallback material only needs to suppress the MD radiation of the NS by burying a certain number of open field lines of the NS. It could not be necessary to screen the whole star and hide all of the surface magnetic field.

After the magnetic filed is buried, it will diffuse back toward the surface through Ohmic diffusion and Hall drift \citep{Thompson1996,Thompson2002,Heyl1998,Beloborodov2009,Gourgouliatos2015,Rogers2017}. The timescales of these processes can be expressed roughly by
 \begin{equation}
t_{\rm Ohm}={4\pi \sigma_c L^2\over c^2}=4.4\times10^6{\rm~ yr}\left({\sigma_c\over 10^{24}{\rm s^{-1}}}\right)^{2}\left({L\over 1{\rm km}}\right)^{2},
\end{equation}
and
 \begin{eqnarray}
t_{\rm Hall}&=&{4\pi n_{\rm e}e^2L^2\over cB}\nonumber\\
&=&6.4\times10^4{\rm~ yr} \left({n_{\rm e}\over 10^{36}~\rm cm^{-3}}\right)\left({L\over 1{\rm km}}\right)^{2}\left({B\over 10^{15}}\right)^{-1},
\end{eqnarray}
respectively, where $\sigma_c$ is the average conductivity of NS matter, $n_{\rm e}$ is the electron density, $L$ is the crust thickness, and $B$ is the hidden magnetic field. By assuming that the surface magnetic field could only be hidden partially, the reemergence timescale of the field could become shorter than those presented above. Nevertheless, it could still be safe to regard the remnant NS of GW170817 as a constant low-field NS during the kilonova and off-axis afterglow emissions. It will be very interesting to detect the astrophysical consequences\footnote{We suspect that the reemergence of the hidden magnetic field from the remnant NSs could have a connection with the phenomena of fast radio bursts \citep{Cao2018}.} of the reemergence of the hidden magnetic field in future.

\subsection{Spin evolution and energy releases}
For a clear illustration (but not demonstration) of the very early evolution of the remnant NS of GW170817, we suppose a NS having an initial spin period $P_{\rm i}=2$ ms\footnote{The initial spin period is taken to be somewhat longer than 1 ms, since the NS at birth could release a great amount of energy to trigger a GRB and to radiate GW due to short but strong stellar fluid instabilities.}, an internal magnetic field $B_{\rm int}=3\times10^{16}$ G that determines an ellipticity $\epsilon=10^{-3}$, and a surface magnetic field of an initial value $B_{\rm p}=10^{15}$ G. The surface filed is further considered to decrease to $B_{\rm p}=6\times10^{11}$ G about a few thousands of seconds later. For simplicity, we assume this $B_{\rm p}$-decline happening abruptly, by according to the GRB afterglow observations. Meanwhile, the internal fields and thus the ellipticity keep constant, i.e., the NS becomes a low-field magnetar. Under these assumptions, we can calculate the spin evolution of the NS by
\begin{equation}
{dP\over dt}={P\over \tau_{\rm gw}}+{P\over \tau_{\rm md}},
\end{equation}
where $\tau_{\rm gw}={5P^4c^5/({512\pi^4GI\epsilon^2})}$ and $\tau_{\rm md}={{6IP^2c^3}/ ({4\pi^2B_{\rm p}^2R_{\rm
s}^6}})$. With a given spin evolution, the luminosity of MD radiation can be derived from
 \begin{equation}
L_{\rm md}(t)={B_{\rm p}(t)^2R_{\rm s}^6\over{6c^3}}\left[{2\pi\over
P(t)}\right]^4.
\end{equation}
At the same time, the luminosity of secular GW radiation is given by
 \begin{equation}
L_{\rm gw}(t)={32GI^2\epsilon^2\over{5c^5}}\left[{2\pi\over
P(t)}\right]^6.
\end{equation}
The calculated results are presented in Figure \ref{NSevolution}. As shown in the right panel, before the decline of the surface magnetic field, the luminosity of MD radiation can be much higher than the GW luminosity, which is benefit for ionizing merger ejecta and energizing GRB afterglow emission. Subsequently, due to the $B_{\rm p}$-decline, the MD radiation is drastically suppressed and the spin-down becomes dominated by the GW radiation.

\section{Summary and conclusions}
The successful joint observations of GW170817, GRB170817A, and kilonova AT2017gfo have indicated the beginning of a new era in
multi-messenger astronomy, by witnessing the first GW radiation from a NS-NS merger, by confirming the origin of short GRBs and the existence of a structured relativistic jet, by identifying the kilonova emission and the origin of {\it r}-process elements, and by helping to probe the physics of the pre- and post-merger NSs.

The existence of a long-lived remnant NS is found to be helpful for understanding the observed properties of the AT2017gfo emission. On the one hand, with an ellipticity of $\sim10^{-4}-10^{-3}$ and a surface magnetic field of $\sim10^{11}-10^{12}$ G, the spin-down of the NS, which is dominated by secular GW radiation, can provide an effective energy source to the merger ejecta to power a delayed kilonova emission component. In this case, we do not need to invoke a fine-tuning structure of the merger ejecta of a very high mass, which is however necessary in the widely-considered multiple opacity model. On the other hand, the intermediate value of the opacity found in this paper can also be naturally accounted for in the presence of the remnant NS, if its surface magnetic filed is initially high. Therefore, a sharp decline of the surface magnetic field is suggested to occur at a few thousands of seconds after the birth of the NS. It is very interesting and important to discover such a particular magnetic field evolution, by observing on-axis the early afterglows of future GW-associated GRBs from NS-NS mergers. In any case, the possible existence of a massive NS after GW170817 would provide a very stringent constraint on the EOS of post-merger NSs.

\acknowledgements
We would like to thank the referee for helpful comments that have allowed us to improve our manuscript.
This work is supported by the National Basic Research Program of China
(973 Program, grant 2014CB845800), the National Key Research and Development
Program of China (grant No. 2017YFA0402600), and the National Natural Science
Foundation of China (grant No. 11473008 and 11573014).  L.D.L. is supported by
the China Scholarship Program to conduct research at UNLV.

\end{document}